\documentclass[10pt,reprint,twocolumn,twoside]{revtex4-1}

\usepackage{graphicx}
\usepackage{amsmath}
\usepackage{bm}

\usepackage{color}

\newcommand{\rev}[1]{{\color{black} #1}}
\begin{document}

\title{Experimental characterisation of nonlocal photon superfluids} 
\author{David Vocke$^{1,*}$, Thomas Roger $^{1}$, Francesco Marino $^{2}$, Ewan M. Wright $^{3,1}$, Iacopo Carusotto $^4$, Matteo Clerici $^1$, Daniele Faccio $^{1,}$}
\email{dev1@hw.ac.uk, d.faccio@hw.ac.uk}
\affiliation{$^1$ Institute of Photonics and Quantum Science, Heriot-Watt University, Edinburgh EH14 4AS, UK \\ 
$^2$ CNR-Istituto Nazionale di Ottica, Firenze, Italy}
\affiliation{$^3$ College of Optical Sciences, University of Arizona, Tucson, USA}
\affiliation{$^4$ INO-CNR BEC Center and Dipartimento di Fisica, Universit\'a di Trento, I-38123 Povo, Italy}

\begin{abstract}
Quantum gases of atoms and exciton-polaritons are nowadays a well established theoretical and experimental tool for fundamental studies of quantum many-body physics and suggest promising applications to quantum computing. Given their technological complexity, it is of paramount interest to devise other systems where such quantum many-body physics can be investigated at a lesser technological expense.
Here we examine a relatively well-known system of laser light propagating through thermo-optical defocusing media: based on a hydrodynamical description of light as a quantum fluid of interacting photons, we \rev{investigate} such systems as a valid, room temperature alternative to atomic or exciton-polariton condensates for studies of many-body physics.
First, we show that by using a technique traditionally used in oceanography, it is possible to perform a direct measurement of the single-particle part of the dispersion relation of the elementary excitations on top of the photon fluid and to detect its global flow. Then, using a pump-and-probe set-up, we investigate the collective nature of low-wavevector sound modes of the fluid and \rev{observe signatures of superfluid behaviour.}
\end{abstract}

\maketitle

\section*{Introduction}
Quantum gases, i.e. systems in which the thermal de Broglie wavelength is larger than the average particle distance are an ever increasingly important area of theoretical and experimental study with applications
as diverse as quantum computing and quantum simulation of general relativity models.
The most important example of such physics is the Bose-Einstein Condensate where a macroscopic number of bosonic particles at low temperature share the same wavefunction. After early work on liquid Helium, the physics of condensates has been experimentally studied in ultracold atomic gases and, more recently, in exciton-polariton fluids in semiconductor microcavities. Building on these latter studies, an ever growing community is now active on the study of the so-called quantum fluids of light, where the many-photons forming the beam are seen as a gas of interacting particles via the optical nonlinearity of the medium \cite{Carusotto2013}. Among the many hydrodynamic effects that are presently being studied in such fluids of light, we can mention turbulence \cite{Picozzi2014}, where the general physical processes found in a fluid or superfluid can be found in many other systems such as plasmas or in astrophysical systems. Another intriguing direction is so called "analogue gravity" where flowing fluids are used to model gravitational spacetime geometries \cite{Barcelo2011, Novello2002, Faccio2013, Unruh1981}. Waves propagating on top of a flowing medium behave the same way as light waves in a gravitational field. Hence, by tailoring the properties and flow geometry of such a medium, it is possible to mimic gravitational spacetimes in a laboratory experiment.
As an alternative to actual flowing fluids, optical analogues obtained using laser beams propagating through nonlinear self-defocusing media or confined in nonlinear optical cavities have been proposed to create a photon fluid with the desired flow pattern.
In these systems the fluid properties can be controlled through the phase and intensity of the incident optical field as well as by the refractive index profile and the structural boundaries of the medium \cite{Marino2008, Marino2009, Fouxon2010}. For example, a theoretical framework for photon fluids \cite{Carusotto2013, Marino2008, Marino2009, Carusotto2014, Larre2014} has shown that sound-like excitations experience a curved spacetime and hence are promising models for analogue gravity experiments. A full quantum simulation showing evidence for analog Hawking radiation from a black hole horizon in a fluid of light was performed in \cite{Gerace2012} and the first experimental reports of black hole horizons in fluids of light were reported in \cite{Elazar2012, Nguyen2014}.
The effective photon-photon interaction in the nonlinear medium arises from the third-order Kerr nonlinearity \cite{Chiao1999}, i.e. the local change in refractive index proportional to the light intensity $\Delta n = n_2 |E|^2$: a defocusing nonlinearity with $n_2 < 0$ giving repulsive interactions is required to observe a dynamically stable photon fluid. Such a nonlinearity may be obtained through the thermo-optic effect, i.e. an intense laser beam locally heats the medium, which reacts by
decreasing the refractive index proportionally to the deposited laser energy thus giving rise to an effective repulsive photon-photon interaction {\cite{Pomeau1993}}. Here, we study and propose a relatively well known nonlinear medium, i.e. the thermal defocusing medium described above however, revisited within the context of a fluid of light. The aim is to study the extent to which such systems can indeed represent a novel platform for
studying, at room temperature, the properties of quantum gases with an emphasis on (but not limited to) the exciting perspective of reproducing superfluid behaviour, phonon propagation according to the Bogoliubov dispersion, and particle generation from a black hole horizon and in an analogue gravity set-up.
In the propagating geometry considered here, the dynamics of the fluid takes place in the transverse plane (x,y) of a laser beam propagating through a bulk nonlinear medium, so that the propagation coordinate z plays the role of the time coordinate $t$ \cite{Larre2014, Carusotto2014}. 
Hence, by imaging the evolution of the beam profile along the $z$-coordinate, it is possible to detect the spatio-temporal evolution of effective sound modes  in the photon fluid (i.e. small oscillations on the transverse beam profile) and extract information on the dispersion relation of its elementary excitations. Using a technique inspired from oceanography, we obtain information on the high wavevector part of the dispersion where excitations have a single-particle nature and we detect the effect of global flow of the fluid. Then, a pump-and-probe technique is presented, where the probe beam is used to create elementary excitations in the fluid generated by the pump. Information on the elementary excitation dispersion is obtained from the lateral shift of the interference pattern of pump and probe beams: signatures of the {collective nature} of the excitations and of the non-locality of the photon-photon interactions are identified and discussed. \rev{An important result of this study is the experimental verification of the existence of a (phonon) wavelength regime in which, using the terminology introduced by Chiao et al., photon superfluidity can be observed \cite{Chiao1999}.}

{\section*{Theoretical model}}

The propagation of a monochromatic laser beam with a {vacuum} optical wavelength
$\lambda$ in a self-defocusing medium can be described {at steady state} by the non-linear
Schr\"odinger equation (NLSE):
\begin{equation}
\partial_z E=\frac{i}{2k}\nabla^2E-i\frac{kn_2}{n_0}\vert E\vert^2E
\label{eq:nlse}
\end{equation}
Where $z$ is the propagation direction, $k=2\pi n_0/\lambda$ the optical
wave vector and $n_0$ is the linear refractive index of the medium. The {Laplacian term
$\nabla^2E$ with respect to the transverse coordinates $\mathbf{r}(x,y)$} describes the linear diffraction
of the laser beam, where the {second, nonlinear, term proportional to $n_2$ refers to the optical Kerr nonlinearity, i.e. an intensity-dependent refractive index. In this work, we will always take $n_2<0$, that is a 
self-defocusing nonlinearity, which guarantees transverse stability of the beams}. By writing the field $E$ in terms of its amplitude
and phase $E=E_{bg}\exp{(i\phi)}$, equation (\ref{eq:nlse}) can be re-written as two differential equations \cite{madelung}
\begin{eqnarray}
\partial_{{\tau}}\rho+\nabla(\rho \boldsymbol{v})=0 \label{eq:conti} \\
\partial_{{\tau}}\psi+\frac{1}{2}v^2+\frac{c^2n_2}{n_0^3}\rho-\frac{c^2}{2k^2n_0^2}\frac{\nabla^2\sqrt{\rho}}{\sqrt{\rho}}=0
\label{eq:euler}	
\end{eqnarray}
{formally identical to the density and phase equations of a two-dimensional BEC, the temporal variable $\tau=zn_0/c$ being proportional to the propagation distance.}
The optical background intensity $\vert E_{bg}\vert^2$ is identified
as the photon fluid density $\rho$ and the phase $\phi$ defines a fluid velocity
$\boldsymbol{v}=(c/kn_0)\boldsymbol{\nabla}\phi=\boldsymbol{\nabla}\psi$.

{The small amplitude perturbations $E=E_{bg}+\epsilon$ ($E_{bg}$ is the background field) can be described within the Bogoliubov theory in terms of sound waves on top of the photon fluid~\cite{Carusotto2014,BECbook}. 
In the spatially homogeneous case where $E_{bg}$ does not depend on the transverse coordinate $\mathbf{r}$, a sound mode of wavevector $\boldsymbol{K}$ and amplitude $\alpha_{\boldsymbol{K}}$ has a plane-wave form
\begin{equation}
\epsilon=\alpha_{\boldsymbol{K}} u_{\boldsymbol{K}}\,
\exp{(-i\Omega \tau+i\boldsymbol{Kr})}+\alpha^*_{\boldsymbol{K}}\,v_{\boldsymbol{K}}
\exp{(i\Omega \tau-i\boldsymbol{Kr})}
\end{equation}
and its angular frequency $\Omega$ (in the temporal $\tau$ variable) satisfies the dispersion relation
\begin{equation}
	(\Omega-\boldsymbol{vK})^2=\frac{c^2 n_2\vert
	E_{bg}\vert^2}{n_0^3}K^2+\frac{c^2}{4k^2n_0^2}K^4.
	\label{eq:bogodr}
\end{equation}
}
Here, $c$ is the speed of light and $v$ the background flow velocity.
{Using the terminology of hydrodynamics,} we note that for low
frequencies the dispersion of the sound modes {has a sonic character $\Omega\propto
K$,} whereas for higher frequencies the second term dominates and the sound
modes {follow a single-particle quadratic dispersion $\Omega\propto K^2$}. The separation
between these two regimes {defines} a characteristic length
$\xi=\lambda/2\sqrt{n_0 \vert n_2\vert\,\vert E_{bg}\vert^2}$, {usually called the {\em healing length} in the BEC literature~\cite{BECbook}. As a result, only long-wavelength sound modes with $\Lambda\gg\xi$ follow a sonic dispersion with a constant sound speed
$c_s=\sqrt{c^2 \vert n_2\vert\,\vert E_{bg}\vert^2/n_0^3}$.}
This is of particular relevance, since only a linear dispersion guarantees
superfluid behaviour~\cite{Carusotto2014} and, in the context of analogue gravity, Lorentz invariance
in the acoustic metric \cite{Marino2008, Macher2009}.\\

{Translating these concepts in to the language of  optics~\cite{Carusotto2014}, $K=\sqrt{K_x^2+K_y^2}$ is the magnitude of the transverse wave vector of the light field and $\Omega=(c/n_0)\Delta K_z$ is related to the change $\Delta k_z$ in the wave vector along the propagation axis.}
To experimentally access the low frequency modes i.e. the $\Lambda\gg\xi$ regime,
the light has to propagate in the nonlinear medium for at least one oscillation period $T$, so with
$\Lambda=c_sT\gg\xi$, one arrives at a minimum propagation length
given by $z_{min}\approx cT/n_0\gg \lambda/\Delta n$. We therefore see  that
for a reasonable nonlinearity of $\Delta n=10^{-4}$, long samples with at least
$10-20$ cm length are required to observe the propagation of modes that lie in the linear regime of the
dispersion. 

\section*{Non-local media}
A natural choice for the nonlinear material is therefore a liquid that exhibits
a thermo-optic nonlinearity. The main advantage is that samples of the
required dimensions can easily be built and that they are isotropic. In thermo-optical
media, self-defocusing of the laser beam arises due to partial absorption of the  beam, which results in heating
effects and thus to a decrease in the refractive index \cite{Carter1984,Sinha2000}.
Thermal nonlinearity is usually highly nonlocal, i.e. the change in
refractive index at any position depends not only on the local intensity,
but also on surrounding field intensities \cite{Krolikowski2001, Minovich2007}.
This is a result of heat conduction inside the nonlinear medium and has
a striking influence on the wave dynamics in nonlinear optics
\cite{Wan2007,Barsi2007a,Ghofraniha2007}. \rev{The medium response may be}
 described by a
response function {$R(\mathbf{r},z)$} so that $\Delta n$ can be written as
\begin{equation} 
	\Delta n(\mathbf{r},z)=\gamma\int R(\mathbf{r}-\mathbf{r}')I(\mathbf{r}')\,d\mathbf{r}'\,dz',
	\label{eq:nonlocalnl}
\end{equation}
with $\gamma$ being a normalisation factor and $I$ the field
intensity.
In general, $R$ depends on the nonlocal process in the material and can be of different shapes. \rev{It is also important to note that due to the long-range nature of thermal diffusion, $R$ will also depend on the boundary conditions, e.g. on the shape and materials used for the cell containing the liquid \cite{Minovich2007}.}
For thermo-optical media an isotropic exponential decay was
proposed for $R(\mathbf{r},z)=1/(2\sigma_L)\exp{(-\sqrt{r^2+z^2}/\sigma_L)}$, where the width of the
response function $\sigma_L$ defines the degree of nonlocality
\cite{Bar-Ad2013,Minovich2007}.\\

\begin{figure}[t!]
 \begin{centering}
	\includegraphics[width=0.48\textwidth]{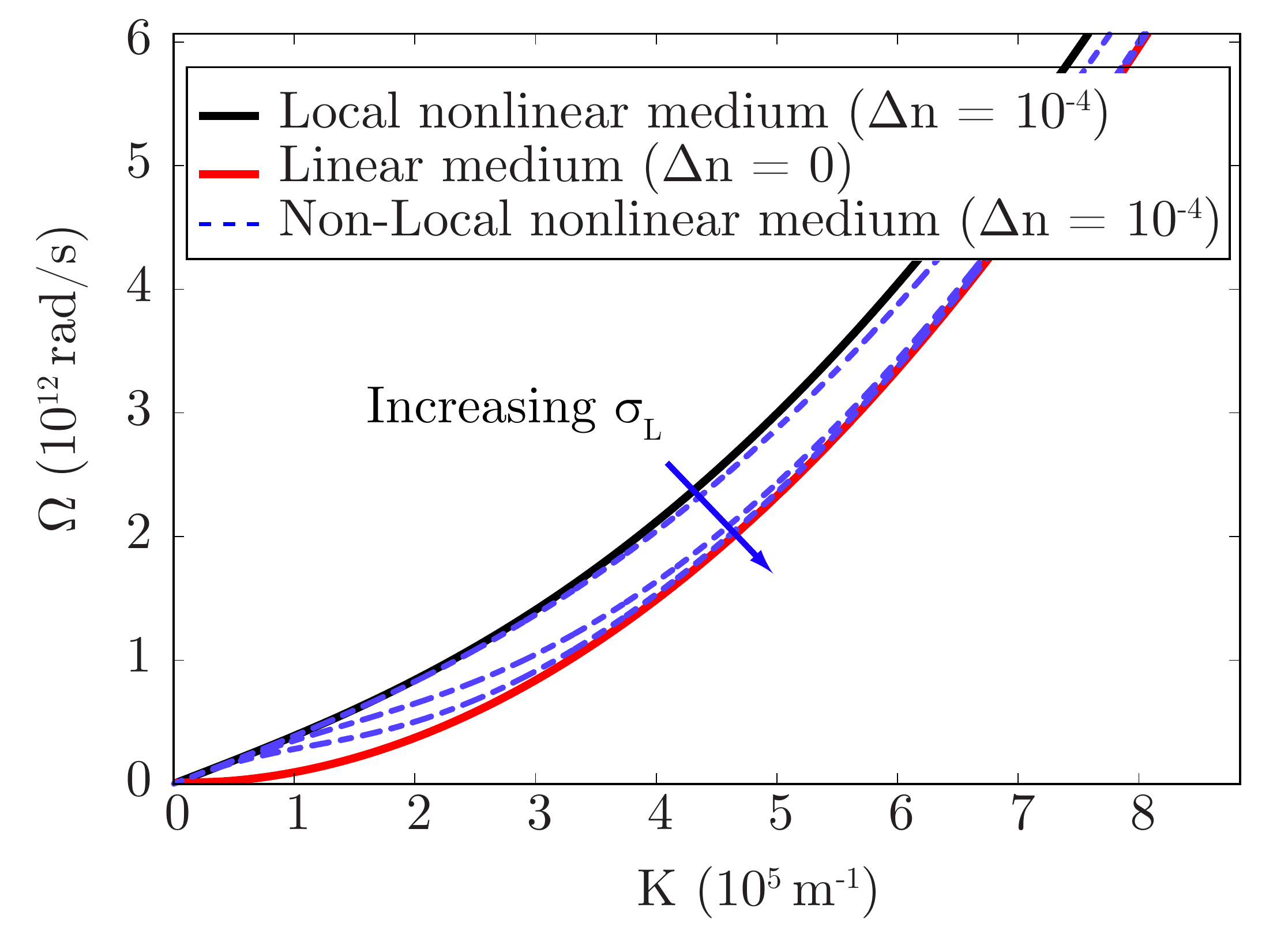}
	\caption{Bogoliubov dispersion relation according to
	Eq.(\ref{eq:bogodr} and \ref{eq:bogodr_nl}). Local nonlinear media ($\Delta n=10^{-4}$) show a linear part
	for wavelengths $\Lambda\gg\xi$ and $|\Delta n|>0$ (black solid line, \rev{$\xi\sim350$ $\mu$m, equivalent to K$=2.7\times10^5$ 1/m}).
	In nonlocal media the \rev{effectiveness of the} nonlinearity is suppressed for wavelengths
	$\Lambda<\sigma_L$ that results in a parabolic  dispersion relation
	(the dotted blue lines correspond to a nonlocal nonlinear dispersion with
	$\sigma_L=1$, $5$ and $10$ $\mu$m).}
	\label{fig:nldr}
	\end{centering}
\end{figure}

{Straightforward extension of the Bogoliubov theory to non-local media leads to a modified Bogoliubov dispersion of the form
\cite{Bar-Ad2013,Pomeau1993}:
\begin{equation}
	(\Omega-\boldsymbol{vK})^2=\frac{c^2n_2\vert
	E_{bg}\vert^2}{n_0^3}\hat{R}(K,n_0\Omega/c)K^2+\frac{c^2}{4k^2n_0^2}K^4
	\label{eq:bogodr_nl}
\end{equation} 
where $\hat{R}$ is the three-dimensional Fourier transform of the response function $R(r,z)$. It is easy to see that well within the paraxial approximation $K/k, v/c \ll 1$, the rescaled Bogoliubov frequency $n_0 \Omega /c \ll K$, so the main contribution to the non-locality comes from the $K$ dependence of $\hat{R}$, and we can simplify $\hat{R}(K,n_0\Omega/c)\simeq \hat{R}(K,0)=1/(1+K^2\sigma_L^2)$ in Eq.~(\ref{eq:bogodr_nl}). \\

Examples of the Bogoliubov dispersion in non-local media with different parameters are shown in Fig.~\ref{fig:nldr}. For very low wavevectors $K\ll 1/\xi,1/\sigma_L$ the nonlinearity is responsible for the linear, sonic shape of the dispersion $\Omega\simeq c_s K$. At very large wavevectors $K\gg 1/\xi,1/\sigma_L$, the dispersion recovers the quadratic single-particle dispersion $\Omega\propto K^2$. The effect of the non-locality is most interesting for $\sigma_L\gg \xi$: in this case, there is a wide range of $1/\sigma_L< K < 1/\xi$ where the nonlocality \rev{suppresses the effectiveness of the} nonlinearity and the dispersion is pushed towards the single-particle dispersion already at $K\simeq 1/\sigma_L \ll 1/\xi$. For a sufficiently small $\sigma_L$ and a sufficiently fast decrease of $R$ (faster than the one we are considering here, e.g. Gaussian), a ring of roton-like minima appears in the dispersion for finite $K$  \cite{Pomeau1993}. As a sonic shape of the dispersion of excitations is essential for gravitational or superfluid analogies to be rigorous, experimental studies in this direction require having control over the nonlinear and nonlocal properties of the material.}\\

In the following, we report experimental evidence of a photon fluid in a nonlocal,
self-defocusing medium. We adapted a method from oceanography to measure the
spatiotemporal dispersion relation and show that by controlling the phase and
intensity of the laser beam, we can imprint a flow as well as tune the velocity
of acoustic waves. We then demonstrate a {second} method, based on tracking the sound
speed of waves in the photon fluid as a function of input power, that allows us to measure both the
local nonlinear refractive index change, $\Delta n$ and the nonlocal length of the medium, $\sigma_L$. \\

{\section*{Experimental results I: oceanographic technique}}

\begin{figure*}[t!]
 	\centering
 		\includegraphics[width=\textwidth]{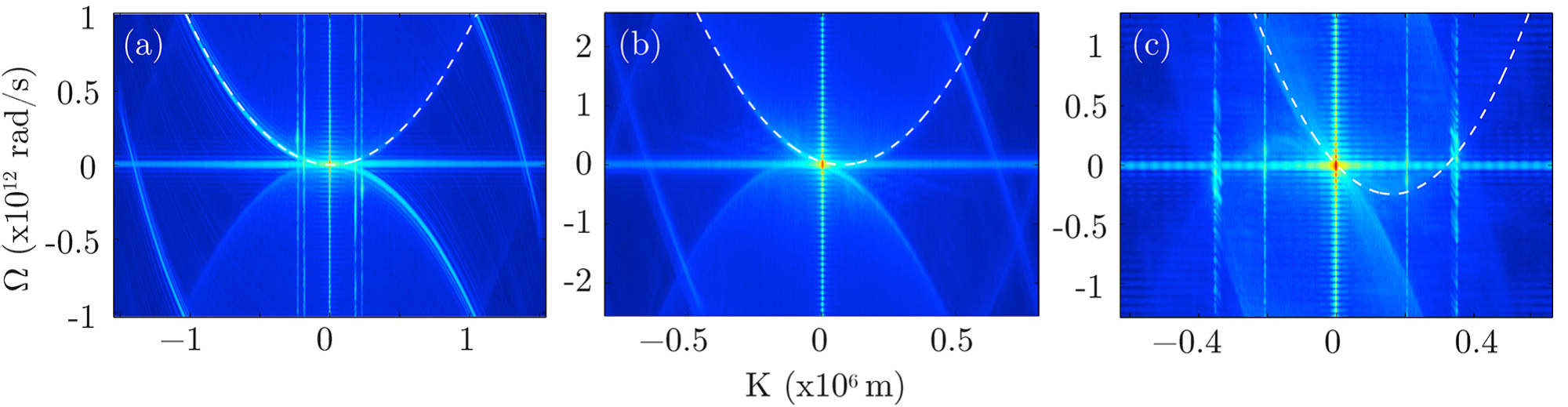}
	
 		\caption{Photon fluid dispersion relation with a scanning distance of
 		1.5 cm, 5 cm, 12 cm (a)-(c).
 		The flow $v$ of the effective medium can be controlled by the phase of
 		the background field ( a) $v_{bg}=0$ m/s, b) $v_{bg}=1.3\times 
 		10^{6}$ m/s c) $v_{bg}=3.0\times 10^{6}$ m/s.}
 		
 		\label{fig:dispersionrelations}
 \end{figure*}

The experiments were carried out by launching a collimated CW-laser beam with 
$532$ nm wavelength through a cylindrical tube with length $L=13$ cm (2 cm diameter) filled with a
methanol/graphene solution. The beam is magnified to a waist diameter of
$1.6$ cm by a 4-f telescope and the central portion with a radius of 1 cm is selected with an aperture to ensure a relatively flat
intensity profile. Methanol is known to have a negative thermo-optic coefficient
of $dn/dT\approx-4\times 10^{-4}$ 1/K but has a very low absorption
coefficient of $\alpha=0.0006$ $cm^{-1}$ \cite{Rindorf2008}. Nanometric graphene
flakes (7 nm average size) are therefore dissolved in the medium in order to
increase absorption. Differently from widely used dyes, graphene does not
exhibit any efficient fluorescence mechanism so that most of the absorbed laser
energy is converted directly into heat. By using a very dilute solution of graphene, it
was possible to prepare samples with a total absorption that could be controllably chosen in the full 0 to $\sim100$\% region over the sample length and adjusted in our experiments so as to have 20\% absorption. A
camera/lens system is mounted on a computer-controlled, longitudinal translation
stage, so that the transverse beam profile can be imaged at different positions
along the propagation axis of the beam.
The imaging was performed at a constant image distance to ensure uniform scaling
in transverse and z-directions during scanning. Since the z-axis maps into a
time coordinate, it is possible to measure the spatiotemporal evolution of small
intensity fluctuations on top of the beam by scanning along z and measuring
the 2D beam profile at defined increments of $\Delta z$. The input beam diameter
was large enough so that linear diffraction is negligible along the sample. To directly measure the dispersion relation, a
technique adapted from oceanographic studies was used where surface waves that naturally
occur were recorded in both space and time and by calculating the Fourier transform one obtains the dispersion of the medium \cite{Stilwell1969, Dugan1996, Weinfurtner2011}. \\ 
\begin{figure*}
	\centering	
	\includegraphics[width=0.9\textwidth]{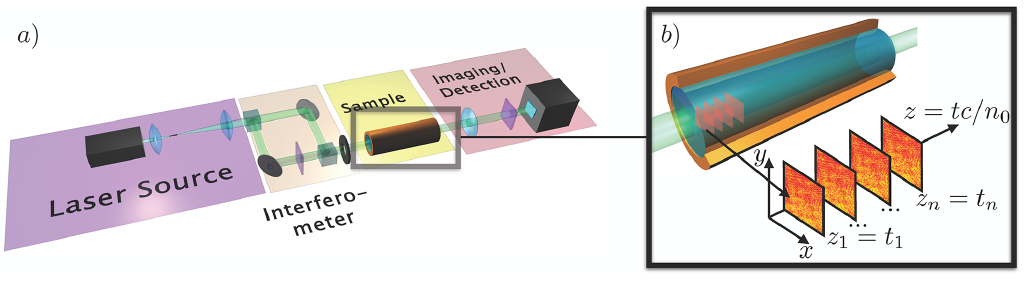}
	\caption{a) Experimental layout used for measuring the dispersion relation.
	A collimated, flat beam is launched through a cylindrical sample filled with a
	methanol/graphene solution and is imaged by a camera/lens system that can be
	translated along the propagation direction of the beam. b) Detail of nonlinear sample showing example beam profiles at different propagation distances (i.e. equivalently for different propagation times) as measured
	in the experiment. \rev{An interferometer placed before the sample generates a pump and probe beams with a controllable relative angle (i.e. wavelength of the photon fluid excitations).}}
	\label{fig:setup}
\end{figure*}
In our experiment, the set of images are stored in a 3D (x,y,z) dataset: the origin of the observed fluctuations is due to \rev{(static)} classical amplitude noise that is always present on the transverse profile of any laser beam (unless one invests significant effort to completely spatially filter it out). The dispersion $I(K_x , \Delta K_z)$ can be calculated via the Fourier transform of a lineout of the measured intensity signal \rev{$I(x, 0, z)$}. 
The resulting spectrum can be converted to frequency using $\Omega= (c/n_0 )\Delta K_z$. To enhance the signal-to-noise ratio, the data was averaged over 200 line-outs along the y direction. Examples of the measured dispersion of
the photon fluid are shown in Figs.~\ref{fig:dispersionrelations}(a), (b) and (c). 
\rev{The spectra show some ``folding'' determined by the lattice of points at which the signal is sampled and then Fourier-transformed. The large signal at ($K=0,\Omega=0$) corresponds to the background pump beam. Some of the spectra also show vertical stripes or other features that do not lie on the dispersion curves and which we were able to attribute to small defects present e.g. on the input or output window facets.}
The main finding of these measurements is the observation of an apparently purely parabolic dispersion for all spatial K-vectors and for all sets of
parameters.
Three different measurements are shown in Fig.~\ref{fig:dispersionrelations}  for three different angles of the
input beam with respect to the sample (and imaging) axis in the $x-z$ plane. This tilt angle
\rev{creates a linear phase gradient along the tilt direction and thus, according to the hydrodynamical equations \eqref{eq:conti} and \eqref{eq:euler} controls the transverse flow
of the photon fluid}.
This can clearly be seen in Figs.~\ref{fig:dispersionrelations}(b) and (c) for which the increasing input angles lead to increasing flow. The data is consistent with Eq.~(\ref{eq:bogodr}) with $\Delta n=0$ and  flow speeds $v_{bg}=$ 0,
1.3, 3.0 x $10^{6}$ m/s [(a), (b) and (c), respectively]. 
We attribute the purely parabolic dispersion  to the nonlocal nature of the
material that, as discussed above reduces the \rev{effectiveness of the} nonlinearity for features smaller than the $\sigma_L$.
A nonlocal length for methanol of $\sigma_L\approx 300$ $\mu$m has been reported
\cite{Minovich2007}, implying that only modes with a much smaller $K<10^4$ m$^{-1}$ would appear to have a linear dispersion (Fig.~\ref{fig:nldr}). \\
We underline that in these measurements we are imaging the various photon
fluid planes inside the actual nonlinear medium. This may cause some issues
related to the fact that we can no longer consider this a linear imaging
measurement and the nonlinearity may lead to deformations of the
various object planes \cite{Barsi2009}. We did however, experimentally verify that objects
 placed at the cell input could be imaged through the sample and
that distortions only appeared to be significant in the presence of sharp
transitions from very low to very high light intensity regions. We therefore
expect that the images of the weak and smooth oscillations lying on top of the
more intense laser beam background are not significantly distorted and that the
overall shape of the measured dispersion curve is indeed correct. This is corroborated by the fact that the
dispersion relation Eq.~(\ref{eq:bogodr}) provides an excellent theoretical fit to the data for the zero-velocity flow case and provides a correct estimate of
the photon fluid flow when the input beam is placed at an angle with respect to
the imaging axis.\\
However, it is clear that a more detailed and precise estimate of the low frequency behaviour and
the presence or absence of superfluid behaviour should be assessed by a
technique that does not require imaging through the nonlinear medium. {The results obtained with such a technique are discussed in the next section.}\\

{\section*{Experimental results II: pump and probe technique}}

\begin{figure}[t!]
		\begin{center}
				\includegraphics[width=0.48\textwidth]{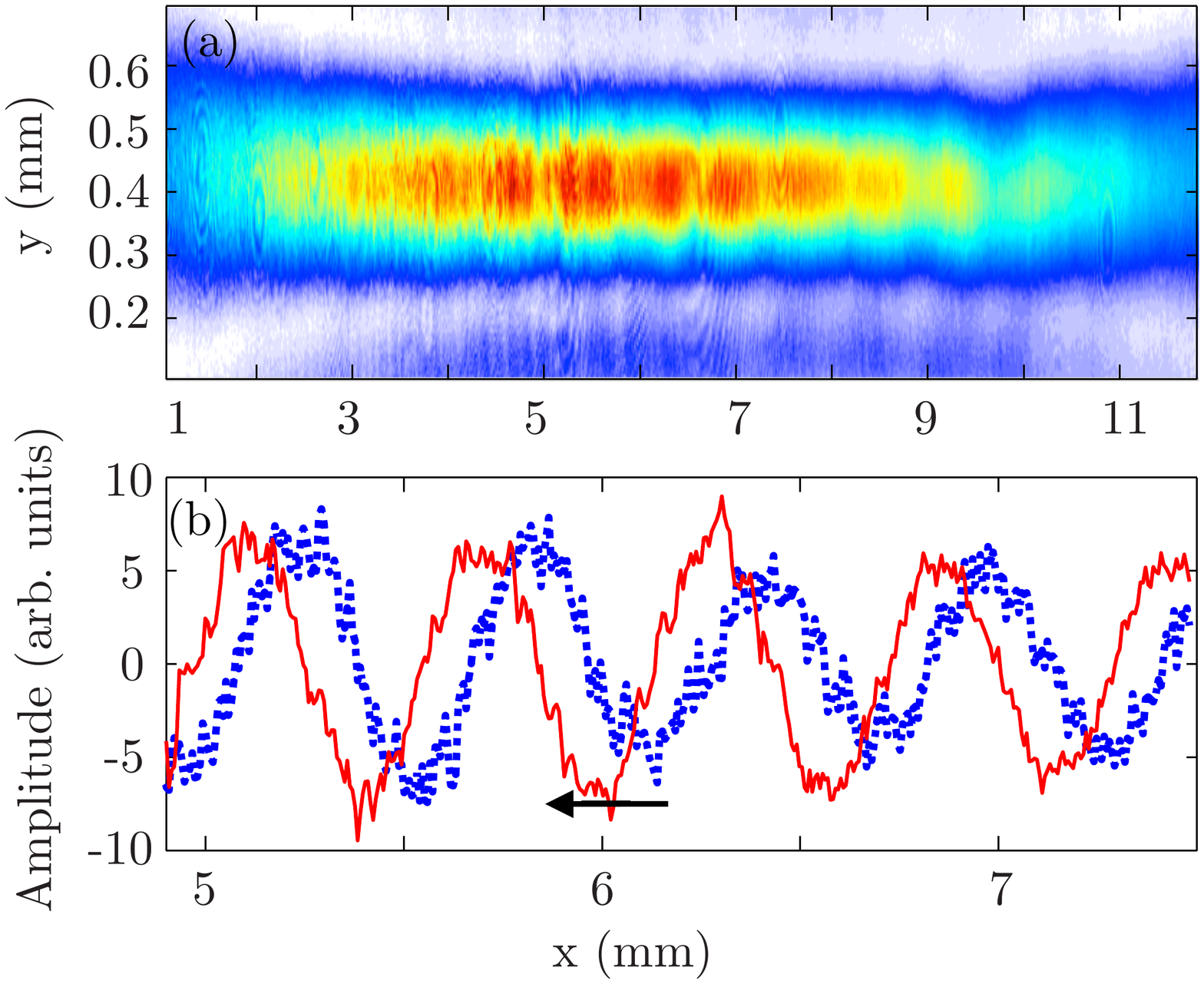}
				\includegraphics[width=0.48\textwidth]{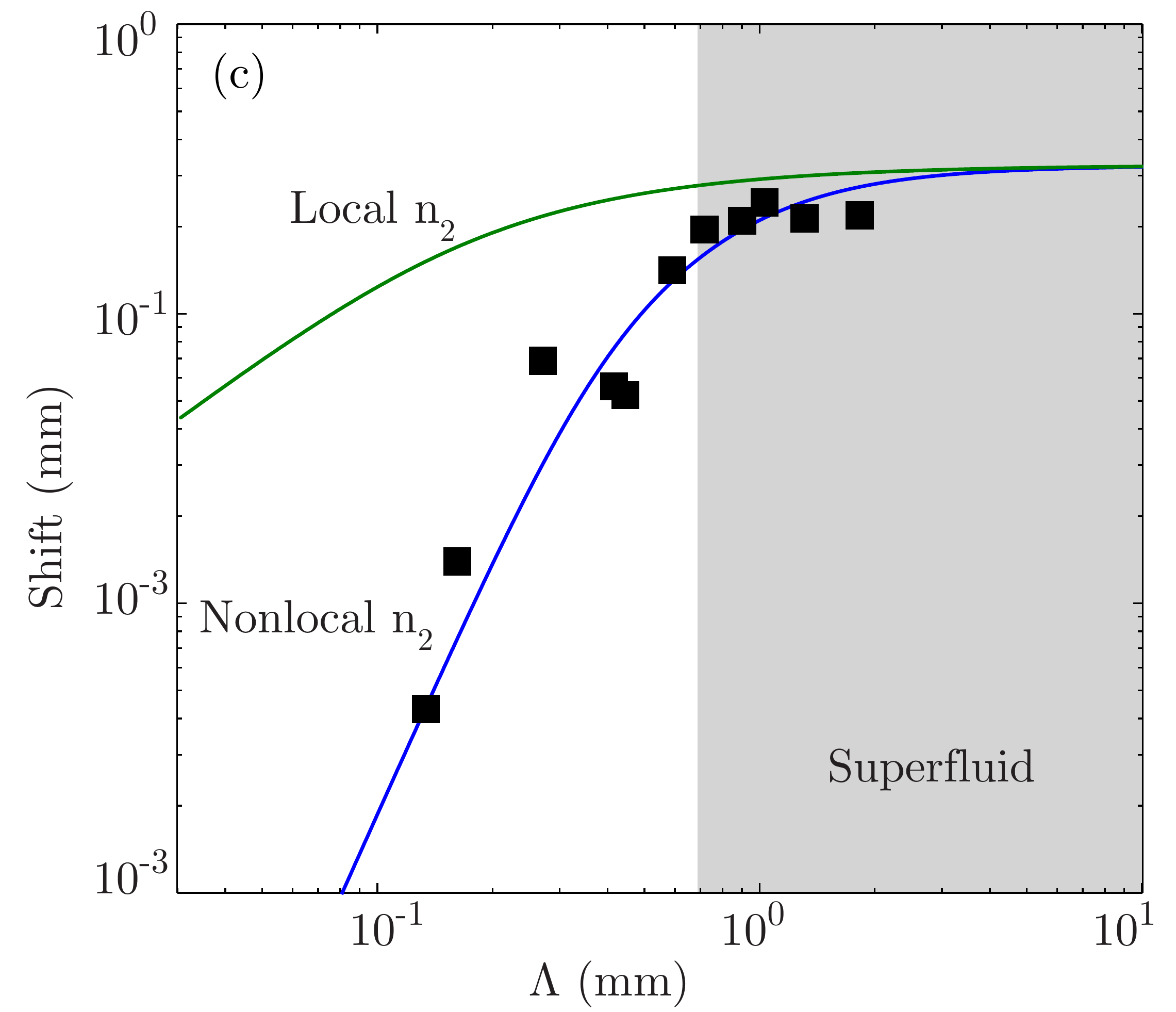}
	\caption{(a) Raw data image of the beam at the sample output. The sound wave is barely visible due to the low (less than 10\%) contrast of the amplitude modulation. (b) Amplitude profile of the sound wave after substracting the pump beam profile for two different input pump powers: low power measurement - dotted blue line and high power measurement (showing a shift in the wave, indicated by the arrow) - solid red line.  (c) Relative shift $\Delta S$ vs.
$\Lambda$: the solid green line  shows the predicted shift for a local
nonlinearity of $\Delta n=-7.6\times 10^{-6}$ - the corresponding healing length is $\zeta=85$ $\mu$m. The solid blue line includes a nonlocal nonlinearity with $\Delta n=-7.6\times 10^{-6}$ and
nonlocal length $\sigma_L=110$ $\mu$m. The shaded gray area ($\Lambda\gtrsim10\zeta$) highlights the region in which, even in the presence of nonlocality, the medium acts as a superfluid. The black squares indicate the measured shift values at various wavelengths, several of which (i.e. for $\Lambda>0.7$ mm) lie in the superfluid region.}
\label{fig:shift}
\end{center}
\end{figure}

{In this experiment, a pump-probe configuration was used to imprint a specific
wavelength on top of the background beam as proposed in~\cite{Carusotto2014}. We} then study the wave dynamics for different
background intensities. The probe beam {was extracted from the pump beam using a beamsplitter and was   attenuated} with \rev{a half-wave plate and a polarising beam splitter}. The probe beam is then
recombined with the pump beam inside the sample: by controlling the relative angle of the two beams we create an interference pattern of the
desired modulation depth ($\approx 5\%$) and {relative wavevector $K$}. \rev{Both the pump and probe beams are loosely focused into the sample using a cylindrical lens pair to create an elliptical beam with a minor axis diameter of $200$ $\mu$m and a Rayleigh range longer than the sample ($\approx20$ cm). } The interference
wave pattern (i.e. the photon fluid sound wave) is finally imaged at the output facet of the sample and a shift
of this pattern along the x  direction can be measured as a function of the
laser power [Fig.~\ref{fig:shift}(a)]. \rev{The final imaging stage is performed with a 1-to-1 imaging f=300 mm lens. In the focal plane we place a mask  to eliminate the four-wave mixed wave, thus suppressing interference between counter-propagating Bogoliubov modes and facilitating the measurement of the phase shift. so as to isolate the pump and probe beams. Our pump-probe setup is similar to that used in the past to study periodic soliton formation \cite{stegeman1,stegeman2} albeit with a lower contrast in the pump-probe interference  pattern and, as we discuss below, our attention is focused on the dynamics of the photon-fluid sound waves.}\\
As the beam propagates through the
nonlinear medium, the phase velocity of the sound wave is determined by $v_{P}=\Omega/K$ with $\Omega$
given by Eq.~(\ref{eq:bogodr}) and hence is a function of the nonlinearity $\Delta n$.
As the pump beam power is increased we therefore expect to observe {a spatial shift of the sound wave pattern by $\Delta S$} due to the increase of the phase velocity.
So by using Eq.~(\ref{eq:bogodr}), we obtain an expression for $\Delta S=\Delta
S(\Delta n,K)$:

 \begin{eqnarray}
 	\Delta S{=\frac{K}{2k}\left[\sqrt{1+\frac{|\Delta n|}{n_0}\,\hat{R}(K)\left(\frac{2k}{K}\right)^2}-1 \right]\,z} 
   \label{eq:Shift2}
 \end{eqnarray}

By measuring $\Delta S$ {for different values of the pump power} we may therefore use Eq.~(\ref{eq:Shift2}) to  estimate both $\Delta n$ and
$\sigma_L$. {As a function of the sound wave wavevector $K$, one can recognize the finite limit $\Delta S \to \sqrt{\Delta n/n_0}$ determined by the speed of sound in the fluid of light. The local vs. non-local nature of the nonlinearity is visible for short wavelengths. In the local case, the phase shift tends to $0$ linearly in $1/K$ with a slope proportional to the refractive index change $\vert \Delta n\vert$. In the non-local case, an extra decay is introduced by the $\hat{R}(K)$ factor: for the case in which which $\hat{R}(K)=1/(1+(\sigma_LK)^2)$, for low $K$ this tends to zero as $K^{-2}$. As a result, the phase shift tends to $0$ much more quickly. While a non-zero $\Delta S$ is by itself a consequence of interactions, the nonlocal effect on the Bogoliubov dispersion is visible in the deviation of $\Delta S$ from a linear dependence on $1/K$.}\\

The experimental data of the relative shifts of several transverse wavelengths $\Lambda=2\pi/K$ between
$120$ $\mu$m and $1.8$ mm at a fixed laser power of 28 mW (fluence, 1 W/cm$^2$) is shown in Fig.~\ref{fig:shift} (solid black squares). \rev{For comparison, the predicted shift in a medium with a purely local ($\hat{R}=1$ ) nonlinearity ($\Delta n=-7.6\times 10^{-6}$) according to Eq.~(\ref{eq:Shift2}) is also plotted  (solid green line). In general, we found that it is not possible to obtain a satisfactory agreement with the experimental trend for any value of $\Delta n$ and a purely local nonlinearity.\\
{On the other hand, if} we take into account the nonlocal factor $\hat{R}(K)=1/(1+K^2\sigma_L^2)$ in Eq.~(\ref{eq:Shift2}), we obtain a very good agreement with the data with $\Delta n=-7.6\times 10^{-6}$ and $\sigma_L=110$ $\mu$m (solid blue line). \\
A remarkable feature of these measurements is that for the longer wavelengths, i.e. for $\Lambda>0.7$ mm the observed shift appears to saturate and, within an 8\% margin, is the same for all wavelengths. In other words, the phase velocity of the waves, $c_s=\Delta S/(Ln_0/c)=3.8\times 10^{5}$ m/s, does not depend on wavelength in this region (shaded in gray in the figure) and the dispersion relation is dominated by the linear behaviour of the Bogoliubov term. This is a clear evidence of the linear dispersion of collective excitations typical of a superfluid or, in the language of general relativity, of Lorentz invariant propagation that is crucial for future analogue gravity experiments. Naturally, the the observation of superfluid features in a system at room-temperature is in itself of great interest and future experiments will be directed at deepening our understanding and control over the flow and wave propagation in this nonlocal photon superfluid.}\\
{A crucial point here is that a standard Z-scan measurement would not be adequate for characterising such a nonlocal nonlinearity. Z-scan relies on translating the sample through the focus of a lens, i.e. through a region in which the transverse length scale of the beam is changing. The nonlinearity analysed 	here shows a clear dependence on the transverse beam dimensions through the $\hat{R}$ response function. In a Z-scan measurement the material would  therefore respond with different effective nonlinearities at different measurement positions, resulting in a distorted Z-scan trace. We explicitly verified that this is indeed the case (data not shown).}\\

\section*{Conclusions}

We have reported a nonlinear optical experiment that aims at characterizing the physics of a photon fluid realized in a cavity-less nonlinear medium with a nonlocal optical nonlinearity of thermal origin.
With a technique used in hydrodynamics, we measured the parabolic dispersion relation for small amplitude, small wavelength noise excitations and highlighted the evidence of a hydrodynamic flow in the presence of a linear phase term on the transverse beam profile.
With a pump-probe technique we investigated the larger wavelength region by measuring the change in the phase velocity of collective excitations as a function of the pump intensity. In this way, we estimated the thermal nonlinearity to $\Delta n=-7.6 \times 10^{-6}$ and the nonlocal length $\sigma_L = 110$ $\mu$m. \rev{This measurement technique also highlighted a superfluid behaviour of the photon fluid, which therefore holds promise both for future room-temperature studies of superfluidity and for analogue gravity measurements that require a Lorentz invariant (i.e. dispersionless) flowing medium.  
Our measurements indicate that there is a readily attainable wavelength region in which such measurements may be feasible in the near future.  However, the sound speed in the superfluid regime demonstrated is relatively low  $c_s=3.8\times 10^{5}$ m/s and will require very long samples in order to observe significant wave propagation. Future work will therefore be directed at increasing the sound speed, e.g. by use of materials with higher nonlinearities. On the other hand, the degree of nonlocality can also be tuned by modifying the sample geometry \cite{Minovich2007} or absorption \cite{Ghofraniha2012} and this opens opportunities for performing related experiments over a broader wavelength range. A crucial and natural demonstration of superfluidity will involve the study of the transition from laminal flow arond an obstacle at subsonic flow speeds to turbulent dissipative flow with the formation of quantised vortices at supersonic flow speeds \cite{pom,Carusotto2014}.  Furthermore, such media show promise for studies of hydrodynamical turbulence in the presence of nonlocal interactions and the study of collective dynamics of incoherent waves. Finally, extension of our experimental apparatus beyond the CW measurements shown here so as to measure also time-dependent fluctuations could provide access to quantum fluctuation features \cite{Larre2014} such as analogue Hawking radiation \cite{Barcelo2011}.}\\

\section*{Acknowledgements}
DF acknowledges financial support from the Engineering and Physical Sciences Research Council EPSRC, grant no. EP/J00443X/1 and from the European Research Council under the European Union's Seventh Framework Programme (FP/2007-2013)/ERC grant agreement n.306559. IC acknowledges financial support by the ERC through the QGBE grant and by the Autonomous Province of Trento, partly through the project "On silicon chip quantum optics for quantum computing and secure communications" ("SiQuro").\\

\end{document}